\documentclass[a4paper,12pt,english]{article}
\usepackage{amsfonts,bm,amssymb,euscript,array,babel}
\usepackage{amsthm}
\usepackage{mathtools}
\usepackage{tikz-cd}
\usepackage{amsmath}
\usepackage{hhline}
\usepackage[amsmath]{empheq}
\usepackage{hyperref}
\usepackage{cancel}
\usepackage{empheq}
\usepackage{fancybox}
\usepackage{mathrsfs}
\usepackage{pdfsync}
\usepackage{tikz-cd}

\newsavebox{\mysaveboxM}
\newsavebox{\mysaveboxT}

\makeatletter

\newcommand{\bbm}{\left(\begin{matrix}}
	\newcommand{\ebm}{\end{matrix}\right)}
\newcommand{\beq}{\begin{eqnarray}}
	\newcommand{\eeq}{\end{eqnarray}}

\makeatother

\newcommand{\be}{\begin{equation}}
	\newcommand{\ee}{\end{equation}}

\newcommand{\beqa}{\begin{eqnarray}}
	\newcommand{\eeqa}{\end{eqnarray}} 
 \def \bea{\begin{eqnarray}} \def\eea{\end{eqnarray}}

\newcommand{\barr}{\begin{array}}
	\newcommand{\earr}{\end{array}}
\numberwithin{equation}{section}

 \def\one{\mbox{1 \kern-.59em {\rm l}}}

\def\bit{\begin{itemize}} \def\eit{\end{itemize}}

\def\({\left(} \def\){\right)}

 \allowdisplaybreaks[3]

\begin{document}
\renewcommand{\title}[1]{\vspace{10mm}\noindent{\Large{\bf
			
			#1}}\vspace{8mm}} \newcommand{\authors}[1]{\noindent{\large
		
		#1}\vspace{5mm}} \newcommand{\address}[1]{{\itshape #1\vspace{2mm}}}
		
\begin{titlepage}
	
	
	\begin{center}
		
		
		\title{ {\Large Generalized symmetries as \\  homotopy Lie algebras }}
		
		\vskip 3mm
		
		  \authors{ \large
		   Larisa Jonke{\footnote{larisa@irb.hr}} }
		 
		 \vskip 3mm
		 
		  \address{ Division of Theoretical Physics, Rudjer Bo\v skovi\'c Institute \\ Bijeni\v cka 54, 10000 Zagreb, Croatia \\
		  
		  \vskip 3mm
		  
		   School of Theoretical Physics,  {Dublin Institute for Advanced Studies} \\ {10 Burlington Road}, {Dublin 4},  {Ireland}}
		
		
		\begin{abstract}
			\noindent
			Homotopy Lie algebras are a generalization of differential graded Lie algebras encoding both the kinematics and dynamics of a given field theory. Focusing on kinematics, we show that these algebras provide a natural framework for the description of generalized gauge symmetries using two specific examples. The first example deals with the non-commutative gauge symmetry obtained using Drinfel'd twist of the symmetry Hopf algebra. The homotopy Lie algebra compatible with the twisted gauge symmetry turns out to be the recently proposed braided $L_\infty$-algebra. In the second example we focus on the generalized gauge symmetry of the double field theory. The symmetry includes both diffeomorphisms and gauge transformation and can consistently be defined using a curved $L_\infty$-algebra.
			
		\end{abstract}
		
	\end{center}
	
	\vskip 2cm
	
\end{titlepage}

\section{Introduction}\label{sec1}

Symmetries are indispensable in the construction of physical theories and they have proven important both in the description and for the understanding of the phenomena  observed in nature.  The successful implementation of the concept of gauge symmetries for the development of the Standard Model of particle physics  motivated additional research focused on  two  important  questions. The first question is related to  finding a precise mathematical  description of quantization and renormalization procedures in quantum field theory. The main  difficulty here is the lack of rigorous formulation of renormalization of infinities inherent in the standard quantum field theory calculations.  The second question is how to  apply the lessons from  standard model physics and quantum field theory to more general settings including gravity. The issue we are facing  in these settings is a need to implement a more general notion of  symmetries, like higher gauge symmetries, non-commutative gauge symmetries and dualities, to name a few.  
A possible systematic approach to both problems is based on the  Batalin-Vilkovisky (BV) formalism  developed for quantization of field theory \cite{MH1,MH2,Costello,Gw}. The formalism itself is intimately tied with  homotopy Lie algebras \cite{js1,csft,ls}, as shown by Zwiebach in his seminal work on closed string field theory \cite{csft}. 
 
 Homotopy Lie algebra or $L_\infty$-algebra is a generalization of a differential graded Lie algebra in which the Jacobi  identity holds only up to homotopy. It is defined on a graded vector space $V=\bigoplus_{d\in \mathbb{Z}} V_d$, where each vector space $V_d$ contains a physical quantity of assigned grading degree $d$. If we focus on gauge symmetry alone, we define a space of gauge parameters and a space containing gauge fields. Then we define maps on and between these spaces, satisfying a number of consistency conditions that define a gauge algebra in question. Importantly, one can extend this construction to include all data of a (classical, perturbative) field theory, by adding appropriate spaces for equations of motions, Noether identities, potential anomalies and even generators of global symmetries, see Refs.\cite{olaf,brano,Jurco:2020yyu} for  motivating reviews and additional references. Furthermore, building on the relation to the BV formalism, one can formulate the  quantum homotopy algebra relevant for calculation of loop amplitudes \cite{Jurco:2019yfd}.

In this   review we discuss two examples of  generalized gauge symmetries formulated in terms of homotopy Lie algebras.  For both of these examples we stress what are the new insights obtained and   comment on open questions and further developments. The review is focused on the relevant symmetry structures and it does not include examples of field theory realization  and related quantization and renormalization issues.

The first example we shall discuss is relevant for  the construction of consistent field theory with  non-commutative gauge symmetries. Despite the long history of the topic,  see  Ref.\cite{rr} for an early review,  the quantization of non-commutative gauge theories is still not fully understood. However,  recent   applications of homotopy (Lie) algebras for the constructions and the understanding   of consistent non-commutative deformations in field theory might offer new clues.

Similarly to the BV formalism, the  $L_\infty$-algebra can be used  to construct consistent deformations of a given gauge-invariant, perturbative theory. This idea was applied in "bootstraping"   non-commutative gauge-invariant theories starting from the  star-deformation of the usual gauge algebra \cite{Vlad}.  This is a powerful method for perturbative (in orders of deformation parameter) construction of non-commutative field theories, but it is rather difficult to obtain explicit all order expressions. Thus, it has been suggested that one could use semi-classical approximation based on symplectic embeddings of almost Poisson structures, where one could obtain closed expressions \cite{sk,vv}. 

There exists another type of non-commutative deformation of gauge symmetry,  based on the Drinfel’d twist of the symmetry Hopf algebra \cite{dt}. This approach offered well-defined geometric structure and compatible differential calculus, but the physical interpretation of the twisted symmetry remained unclear,  see e.g. \cite{book}.   Recent attempts to understand twisted non-commutative field theories   in terms of $L_\infty$-algebra structures resulted in the construction of braided $L_\infty$-algebras \cite{Marija,GR} with appropriately adjusted BV formalism \cite{alex}.    In Sect.\ref{sec3}  we discuss some aspects of twisted gauge symmetry in the framework of braided $L_\infty$-algebra. We show that the  $L_\infty$-algebra can be rather naturally extended to a  Hopf algebra  of symmetric graded tensor space \cite{univ}.  Then we  twist this extended  $L_\infty$-algebra with a Drinfel'd twist,  simultaneously twisting its modules.  Taking the $L_\infty$-algebra  as its own (Hopf)  module,  we obtain the  braided $L_\infty$-algebra  constructed in Ref.\cite{Marija}.

The second example originates  from the gauge symmetry of double field theory. Double field theory (DFT) arose as a proposal of a field theory realizing the T-duality symmetry of strings \cite{T,S1,S2,HZ1}. It is a field theory defined on a  doubled $2d$-dimensional configuration space that enjoys global $O(d,d)$  symmetry.  The fields  of the theory, for the purpose of this review\footnote{See e.g.\cite{coimbra} for the geometric description of the dilaton field.},  are  the $d$-dimensional   bein  $e$ and the 2-form Kalb-Ramond field  $B$. The theory also possesses  local, gauge symmetry combining standard diffeomorphisms and gauge transformations of $B$, 
provided that the physical fields satisfy a set of constraints known as the strong constraint. In particular,  the algebra of gauge transformation    defined by the C-bracket, 
closes only for fields and gauge parameters obeying the strong constraint. After imposing  the strong constraint, the C-bracket reduces to the Courant bracket, the properties of which are captured by  Courant algebroids \cite{C90,LWX,Pavol1}.

However, in order to describe the properties of the original C-bracket of the theory one needs a more general structure. The first proposal was given in Ref.\cite{sd} using the geometric structure of a pre-$NQ$ manifold. This structure is defined on non-negatively  graded manifolds with a degree 1 vector field  which does not square to zero, with the obstruction controlled by the strong constraint.  The proposal we shall review here is the DFT algebroid \cite{p1},  which admits a cohomological vector of degree 1 on a graded manifold with general $\mathbb{Z}$-grading \cite{cc}.     In Sect.\ref{sec2} we  discuss the construction of the relevant $Q$-structure based on the  appropriately defined curved $L_\infty$-algebra corresponding to the DFT algebroid \cite{cc,thesis}. This construction represents the first step toward formulation of the relevant world-volume action as proposed in \cite{thesis}.

\section{Hopf algebra \& homotopy Lie algebras}\label{sec3}

An $L_\infty$-algebra  can be  defined in several equivalent ways, and here we follow Ref.\cite{lada} to define it 
 as a ${\mathbb{Z}}$-graded vector space $X=\bigoplus_{d\in \mathbb{Z}} X_d$ with multilinear graded symmetric maps $b_i : X^{\otimes i}\to X$ of degree 1 such that the coderivation $D=\sum_{i=0}b_i$ is nilpotent.    The condition of $D^2=0$   generates the homotopy relations for the maps $b_i$ of the corresponding $L_\infty$-algebra. 
As an example we write  the first few homotopy relations $ \forall x\in X$:
\begin{align}\label{012hr}
    &b_1(b_0)=0~,\\
    &b_2(b_0,x)+b_1^2(x)=0~, \nonumber\\
    &b_3(b_0,x_1,x_2)+b_2(b_1(x_1),x_2)+(-1)^{\lvert x_1\rvert\lvert x_2\rvert}b_2(b_1(x_2),x_1)+b_1(b_2(x_1,x_2))=0~.\nonumber
\end{align}
When $b_0=0$ the $L_\infty$-algebra is called flat and $b_1$ is differential. In that case there exist cochain complex underlying the  flat $L_\infty$-algebra,
\begin{equation*}
    \cdots   \overset{b_1}{\longrightarrow}   X_{i} \overset{b_1}{\longrightarrow}   X_{i+1} \overset{b_1}{\longrightarrow}     \cdots 
\end{equation*}
and  for $b_0\neq 0$ the $L_\infty$-algebra is called curved,  constant element $b_0 \in X_1$ is called curvature. 

 It is important to note that the  structure maps $b_i$ are defined on the whole graded symmetric tensor algebra
 ${\mathbf{S}}(X):=\bigoplus_{n=0}^\infty S^nX$
where $X$ is a $\mathbb{Z}$-graded vector space over the field $K=S^0X$. We denote the degree of a homogeneous element $x_i\in X$   as $\lvert x_i\rvert$, and the graded  symmetric tensor product as $\vee$.  
 The  maps $b_i:S^jX\to S^{j-i+1}X$ act on the full tensor algebra as a coderivation:
\begin{equation}
 \label{mapbi}
b_i(x_1\vee\ldots\vee x_j)=\!\!\!\!\!\sum_{\sigma\in \mathrm{Sh}(i, j-i)}\!\!\!\!\!\epsilon(\sigma;x)b_i(x_{\sigma(1)},\ldots , x_{\sigma(i)})\vee x_{\sigma(i+1)}\vee\ldots\vee x_{\sigma(j)}~, j\geq i~,
\end{equation}
where $\epsilon(\sigma;x)$ is the Koszul sign,
$$x_1\vee \cdots \vee x_k=\epsilon(\sigma;x)x_{\sigma(1)}\vee \cdots \vee x_{\sigma(k)},\qquad x_i\in X~,$$
and  $\mathrm{Sh}(p, m-p)\in S_m$ denotes those permutations ordered as $\sigma(1)<\cdots<\sigma(p)$ and $\sigma(p+1)<\cdots<\sigma(m)$. We use the conventions that Sh$(n,0)=\mathrm{Sh}(0,n)=\mathrm{id} \in S_n$.
Introducing the permutation map $\tau^\sigma:X^{\otimes i}\to X^{\otimes i}$
where the  $\tau^\sigma$ denotes the action of permutations $\sigma$ including the Koszul sign, e.g. the non-identity permutation of two elements is:
$$\tau^\sigma(x_1\vee  x_2)=(-1)^{\lvert x_1\rvert\lvert x_2\rvert}x_2\vee x_1~,$$
we can rewrite the coderivation maps \eqref{mapbi}
\begin{equation*}\label{mapbimap}
b_i\circ\mathrm{id}^{\vee j}=\sum_{\sigma\in \mathrm{Sh}(i, j-i)} (b_i\vee \mathrm{id}^{\vee (j-i)}) \circ \tau^\sigma~,\qquad j\geq i~.
\end{equation*}
Using this notation, the homotopy relations, one for every $i\geq 0$,   can be written in the closed form 
\begin{equation}\label{closed}
\sum_{j=0}^i \sum_{\sigma\in \mathrm{Sh}(j,i)} b_{i-j+1}(b_j\vee \mathrm{id}^{\vee i})\circ\tau^\sigma=0~. \end{equation}
The coderivation  $D=\sum_{i=0}b_i$  satisfies the co-Leibniz property,
\begin{equation}
    \Delta\circ D=(1\otimes D +D\otimes 1)\circ \Delta~,\nonumber
\end{equation}
with the coproduct map $\Delta: {\mathbf{S}}(X)\to {\mathbf{S}}(X)\otimes {\mathbf{S}}(X)$ 
\begin{equation*}\label{map}
 \Delta\circ \mathrm{id}^{\vee m}= \sum_{p=0}^{m}\sum_{\sigma\in \mathrm{Sh}(p, m-p)}(\mathrm{id}^{\vee p}\otimes \mathrm{id}^{\vee(m-p)})\circ \tau^\sigma~,\qquad p,m\geq 0~,
\end{equation*}
defining the  coalgebra structure on ${\mathbf{S}}(X)$.
Thus, the $L_\infty$-algebra can be defined as a coalgebra with coderivation and counit
 $\varepsilon:{\mathbf{S}}(X)\to K$, where $\varepsilon(1)=1$ and $\varepsilon(x)=0,\;x\in X$.
 
 Moreover, the   graded symmetric tensor algebra  ${\mathbf{S}}(X)$ has an algebra structure given by the graded  symmetric tensor product $\vee$ 
and  a unit map $\eta:K\to{\mathbf{S}}(X)$, where $\eta(1)=1$. The algebra and coalgebra structure on ${\mathbf{S}}(X)$ are compatible and make up a bialgebra, which  furthermore admits a graded antipode map  $S$
\begin{equation}
     \label{gS}
S(x_1\vee\cdots\vee x_m)=(-1)^m(-1)^{\sum_{i=2}^m\sum_{j=1}^{i-1}\lvert x_i\rvert\lvert x_j\rvert}x_m\vee\cdots\vee x_1~.\nonumber
\end{equation}
 Thus it is possible to extend the homotopy Lie algebra defined by the coalgebra structure on the graded symmetric tensor space  ${\mathbf{S}}(X)$ to a  cocommutative and coassociative Hopf algebra with compatible coderivation \cite{univ}.  
 
 This observation was used in Ref.\cite{univ} to introduce a non-(co)commutative deformation of a Hopf/$L_\infty$-algebra  in the Drinfel'd twist approach. We twist a Hopf algebra $H$ using a twist element ${\cal F}\in H{\otimes}H$, which is invertible and satisfies:
\begin{align*}
 ({\cal F}\otimes 1)(\Delta\otimes \mathrm{id}){\cal F}&=(1\otimes {\cal F})( \mathrm{id}\otimes \Delta){\cal F}~, \\
(\varepsilon\otimes \mathrm{id}){\cal F}=1\otimes 1&=( \mathrm{id}\otimes\varepsilon){\cal F}~.
\end{align*}
It was shown  that an Abelian twist $\mathcal{F}$ of a Hopf algebra $H$  results in a new Hopf algebra  where only the coproduct gets deformed \cite{Majid,admw}: 
\begin{equation*}
\Delta^{\mathcal{F}}(h)=\mathcal{F}\Delta(h)\mathcal{F}^{-1},\qquad h\in H~.
\end{equation*}
  Thus using the Drinfel'd  twist we obtain a twisted $L_\infty$-algebra with deformed coalgebra sector. In the spirit of deformation quantization we simultaneously twist the Hopf algebra modules. 
  Taking the Hopf algebra\footnote{In lieu with standard notation for Hopf algebra we shall denote this algebra as $(L_\infty, \vee,\Delta,\epsilon,S)$, where $L_\infty$ denotes algebra as a vector space, $\vee$ and $\Delta$ are product and coproduct respectively,  $\epsilon$ is counit and $S$ antipode.} $(L_\infty, \vee,\Delta,\epsilon,S)$ as a module  itself, one  obtains another Hopf  algebra
$(L_\infty, \vee_\star, \Delta_\star,\epsilon,S_\star)$ with:
\begin{align*}
x_1\vee_\star x_2&=\bar f^\alpha(x_1)\vee \bar f_\alpha(x_2)~,\\
\Delta_\star(x)&=x\otimes 1+\bar R^\alpha\otimes \bar R_\alpha(x)~,\\
 S_\star(x)&=-\bar R^\alpha(x) \bar R_\alpha~.
\end{align*}
Here we used Sweedler's summation notation to write the twist element and its inverse
\begin{equation*}
    \mathcal{F}=f^{\alpha}\otimes f_{\alpha},\qquad\mathcal{F}^{-1}=\bar{f}^{\alpha}\otimes\bar{f}_{\alpha}~.
\end{equation*}
The invertible 
 $\cal R$-matrix ${\cal R}\in {\bold S}(X)\otimes {\bold S}(X)$  induced by the twist,
\begin{equation*}
{\cal R}=f_\alpha \bar f^\beta\otimes f^\alpha \bar f_\beta=:R^\alpha\otimes R_\alpha~,\; {\cal R}^{-1}=\bar R^\alpha\otimes \bar R_\alpha\end{equation*}
  controls the non-commutativity of the $\vee_\star$-product
and provides a representation of the permutation group \cite{book}. 
 In particular, 
 the action of a non-identity permutation of two elements is:
 \begin{equation}\label{trr}
\tau^\sigma_R(x_1\vee_\star x_2)=(-1)^{\lvert x_1\rvert\lvert x_2\rvert}\bar R^\alpha (x_2)\vee_\star \bar R_\alpha(x_1)~,\end{equation}
 and it  squares to the identity for triangular\footnote{Abelian twist induces triangular $\cal R$-matrix.} $\cal R$-matrix. 
Using the braided permutation map \eqref{trr} we can write the coproduct   on the  whole tensor algebra:
\begin{equation}
\label{mapstar}
 \Delta_\star\circ \mathrm{id}^{\vee_\star m}= \sum_{\sigma\in \mathrm{Sh}(p, m-p)}(\mathrm{id}^{\vee_\star p}\otimes \mathrm{id}^{\vee_\star(m-p)})\circ \tau^\sigma_R~,\qquad p,m\geq 0~.
    \end{equation}
The   coderivation $D_\star=\sum_{i=0}^\infty b_i^\star$
is defined in terms of braided graded symmetric maps $b_i^\star$:
\begin{align} \label{mapbimapstar}
b_i^\star\circ \mathrm{id}^{\vee_\star j}&=\sum_{\sigma\in \mathrm{Sh}(i, j-i)} (b_i^\star\vee_\star \mathrm{id}^{\vee_\star (j-i)}) \circ \tau^\sigma_R~,\qquad j\geq i~,\\
b_i^\star(\ldots,x_m,x_{m+1},\ldots)&=(-1)^{\lvert x_m\rvert\lvert x_{m+1}\rvert}b_i^\star(\ldots,\bar R^\alpha(x_{m+1}),\bar R_\alpha(x_{m}),\ldots)~,\nonumber
\end{align}
with the condition  $D_\star^2=0$ reproducing the deformed homotopy relations:
\begin{equation}\label{closedstar}
\sum_{j=0}^i \sum_{\sigma\in \mathrm{Sh}(j,i)} b^\star_{i-j+1}(b^\star_j\vee_\star \mathrm{id}^{\vee_\star i})\circ\tau^\sigma_R=0~. \end{equation}
assuming   that the maps $b_i^\star$ commute with the action of the twist generators. 
The braided coproduct \eqref{mapstar} and the compatible  coderivation \eqref{mapbimapstar} equivariant under the action of the degree 0 twist element  reproduce,  in the coalgebra picture,   the braided $L_{\infty}$-algebra constructed in \cite{Marija}.  

Going beyond just the symmetry structure, the  braided gauge symmetry was successfully applied in the construction of field theories, including the braided version of general relativity \cite{Marija},    a braided version of BF theory in arbitrary dimensions  and a new braided version of non-commutative Yang–Mills theory for
arbitrary gauge algebras \cite{GR}. The quantization of the theories with non-commutative braided symmetry is being developed and some preliminary results  can be found in Refs.\cite{alex,cor}.

 \section{DFT algebroid \& homotopy Lie algebras}\label{sec2}

The second example of generalized gauge symmetry we discuss originates in the string-inspired double field theory as described in the Introduction.   The symmetry algebra of the theory  is governed by the C-bracket, whose properties are axiomatically organized in the notion of a DFT algebroid \cite{p1}, in much the same way as the properties of the Courant bracket are collected in the notion of a Courant algebroid.  A DFT algebroid is defined as a rank $2d$ vector bundle $E$ over $2d$-dimensional manifold ${\cal M}$ corresponding to doubled configuration space of DFT. The skew-symmetric bracket on its sections corresponds to a C-bracket, while  the twist of the bracket corresponds to background fluxes of DFT. The pairing on the bundle corresponds to the $O(d,d)$ metric which induces the constant $O(d,d)$ metric on the doubled configuration space.  An anchor map from the sections of the  bundle to the tangent bundle of the base manifold is not a homomorphism of   bundles and, more interestingly, it is invertible \cite{cc}. That is why the anchor  can be related  to the generalized bein of DFT \cite{diego}, where we packaged the field content of the theory, i.e., the d-dimensional bein and Kalb-Ramond field.  

In order to  define a DFT algebroid in terms of curved $L_\infty$-algebra  we start with the following graded vector spaces \cite{cc,thesis}
\[
\begin{array}{ccccccc}
X_{-2}&\oplus  & X_{-1}&\oplus   & X_0 &\oplus  & X_1  \\
f\in C^\infty({\cal M}) & & e\in \Gamma(E) & &  h\in \mathfrak{X}(\mathcal{M})& &{\eta}
\end{array}\]
 Here we present what was called extended $L_\infty$-algebra in Refs.\cite{cc,thesis}, effectively including the base space in the definition. One should think of the  basis of $X_0$ as the one induced by coordinates $x^A$ of a coordinate patch $U\in {\cal M}$
that contains point p such that $x^A\rvert_p = 0$. The vector space $X_1$ is spanned by the constant curvature element $\eta$ of degree 1 given by the induced metric on the target.   As  a next step we define the degree 1 maps $b_i$ such that the homotopy relations \eqref{closed} reproduce the defining properties of the DFT algebroid.  For simplicity, in the following we present just a part of the list of maps, the rest can be found in \cite{cc,thesis}:
\begin{align*}
    & b_0=\eta~, b_1(f)={\cal D}f~,b_1(e)=\rho(e)~, b_2(\eta,f)=-\small{\frac{1}{2}}\eta^{-1}(df)~, 
\end{align*}
where ${\cal D}:C^\infty({\cal M})\to\Gamma(E)$ is the derivative defined through $\langle {\cal D}f,e\rangle=\small{\frac{1}{2}}\, \rho(e)f$, using the $O(d,d)$ pairing on the sections of the bundle and the anchor map $\rho:E\to T{\cal M}$.   Applying the second homotopy relation in \eqref{012hr} to an element in $X_{-2}$, we obtain that it is satisfied provided 
  \begin{equation}\label{a1}
 (\rho\circ {\cal D})f={\small\frac{1}{2}}\eta^{-1}(df)~,     
  \end{equation}
  which is one of the axioms of the DFT algebroid \cite{cc}.

Once we are given the $L_\infty$ structure, we can always construct a cohomological vector $Q$, i.e., we can define the corresponding  $Q$-manifold locally \cite{aksz}. First we determine the structure constants of the algebra by evaluating the  maps $b_i$ on the basis of the graded vector spaces $\{\tau_\alpha\}$
\begin{equation*}
    C^\alpha_{\beta_1\ldots\beta_i}\tau_\alpha=b_i(\tau_{\beta_1},\ldots,\tau_{\beta_i})~.
\end{equation*}
Then we construct the cohomological vector $Q$  
$$Q=\sum_{i=0}^{\infty}\frac{1}{i!}C^\beta_{\alpha_1...{\alpha_i}}z^{\alpha_1}\cdots z^{\alpha_i} \frac{\partial}{\partial z^\beta}~,$$
on the  basis $\{z^{\alpha}\}$ of the dual  graded vector space $X^\star$.
In the case of the DFT algebroid we locally write the Q-manifold as 
$\mathbb{R}[-1]\oplus \mathbb{R}^{2d}\oplus \mathbb{R}^{2d}[1]\oplus \mathbb{R}^{2d}[2]$ with the coordinates\footnote{ Here latin indiced from the beginning of alphabet $\{A,B,...\}$ denote  target indices, while the ones from the middle of alphabet $\{I,J,...\}$ denote bundle indices, just indicating the origin of coordinates on the graded manifold. } $\{s^{AB},x^A,e^I,f_A\}$. Note that the coordinate $s^{AB}$ spans a 1-dimensional vector space. While the full expression for the $Q$ vector is given in Ref.\cite{thesis}, here we set the possible twist of the bracket to zero and anchor map to identity and obtain:
 	\begin{align*}
 	Q&=\eta^{AB}\frac{\partial}{\partial s^{AB}}+\left(\delta^A_Ie^I-\small{\frac{1}{2}}s^{AB}f_B\right)\frac{\partial}{\partial x^A}+\small{\frac{1}{2}}\hat{\eta}^{IJ}\delta^A_If_A\frac{\partial}{\partial e^J},
 \end{align*}
  The constant $O(d,d)$ metric on the bundle is given in the local coordinates as $\hat\eta_{IJ}$, 
  \begin{equation*}\hat\eta_{IJ}=\begin{pmatrix}
	0 & \delta^i_j\\
	\delta^j_i & 0
\end{pmatrix},\qquad i,j=1,\ldots,d~.\label{eq:eta}
\end{equation*}
 and similarly for the  induced constant  $O(d,d)$ metric on the base   denoted as $\eta_{AB}$. In this simplified example, the only condition stemming from the nilpotency of $Q$ is 
  $$\eta^{AB}=\hat\eta^{IJ}\delta^A_I\delta^B_J~,  $$
  which is local coordinate expression of the axiom \eqref{a1}. 
  In comparison with the construction of pre-$NQ$ manifold of Ref.\cite{sd}, we included the induced $O(d,d)$ metric  in the DFT algebroid structure as a curvature element. Thus we obtained an additional coordinate on the $Q$-manifold of negative degree, $s^{AB}$, and bona fide $Q$-vector that is nilpotent.
 
The cohomological vector $Q$ is related to the BRST operator in the Batalin-Vilkovisky formalism where we replace  $z^{\alpha}$  with physical fields, and   even at the level of the classical action it represents an important geometrical structure. Given a graded $Q$-manifold with compatible symplectic structure, known as a $QP$-manifold, one can define a field theory over this graded target space. In terms of $L_\infty$-algebra, the symplectic structure corresponds to  a cyclic pairing compatible with the graded structure making the algebra cyclic $L_\infty$-algebra. Namely, the $L_\infty$ symmetry algebra  can always be extended to include equations of motion, but to write the corresponding action one needs additional structure in the form of compatible cyclic pairing.  The relation between $QP$-manifolds and $L_\infty$-structure is particularly nice in the context of the topological field theories, where one can directly write  the solution of the classical master equation using the $QP$-structure of the  target space \cite{aksz}. However, there are few recently analysed  examples where one does not have compatible symplectic structure at hand, but can, nevertheless construct the solution of a classical master equation. These examples belong to a class of topological sigma models with Wess-Zumino term,  in particular,  twisted Poisson sigma model \cite{tn}, Dirac sigma model \cite{us} and more generally, twisted R-Poisson sigma models \cite{th,tgn}.

In the context of the DFT algebroid, the relevant question is how to construct the classical field theory action functional compatible with the given  symmetry structure. Using the $L_\infty$-algebra structure, one can construct  equations of motion and propose corresponding  world-volume action \cite{thesis}. On the other hand, in the case of odd-dimensional manifolds one can use a contact structure to construct the analog of symplectic potential known as the contact form,  and the hamiltonian vector field (function). Therefore, the   question would be if  the graded Q-manifold underlying the DFT algebroid admits graded contact structure  \cite{Mehtaphd,Mehta,Grabowski} which might be pulled back on the world-volume.

 \section{Outlook}\label{sec4}

In this short review we  argued that homotopy algebras provide  useful framework for describing generalized gauge symmetries. In particular, we showed how to relate an $L_\infty$-algebra with a Hopf algebra, and used this relation to (re)derive the braided $L_\infty$-algebra structure.  In short, the braided $L_\infty$-algebra  is shown to be a twisted Hopf algebra module \cite{Marija,univ}.
Furthermore, we have shown that the generalized gauge symmetry of  double field theory can be described using curved $L_\infty$-algebra, a local structure corresponding to the DFT algebroid. In this context it would be interesting to analyze the relevant global symmetries and their potential 't Hooft anomalies. One question is if the infinitesimal   gauge transformations could be integrated to corresponding finite ones. In the standard case of Lie algebras, it is known due to Lie’s third theorem that the integrated structure corresponds to Lie groups. For Lie algebroids in general it was shown that there are obstructions to integration to higher structures like Lie groupoids, see review \cite{16} for a categorical approach. For homotopy algebroid structures or $L_\infty$-algebroid the analysis was done in \cite{17} using the graded manifold framework.  From physical point of view, the global symmetries are important for understanding of the anomaly structure of the underlying theories. In particular, the investigation of global symmetries  through mixed anomaly lead to identification of 2-group structure underlying some standard Abelian and non-Abelian gauge field theories in four dimensions \cite{19}. Important ingredient in this identification is the existence of a higher (1-form, in nomenclature of \cite{19}) global symmetry with conserved 2-form current. The coupling of this current to a 2-form background gauge field modifies the gauge transformation of the gauge fields by inducing a shift proportional to the field strength of the 1-form gauge field. In general, a $q$-form global symmetry is a global symmetry for which the charged operators are of space-time dimension $q$ (like Wilson lines) and the charged excitations have $q$ spatial dimensions (like membranes) \cite{20}.

Finally, we would like to end this contribution by noting that homotopy algebras can provide additional insight even in the case of standard field theories. One important aspect we haven't commented so far is related to the equivalence relations between homotopy algebras, in particular the $L_\infty$-quasi-isomorphisms. These are  shown to relate equivalent field theories, thus providing an explanation for a number of phenomeologically established recursion relations and dualities  relevant in calculations of scattering amplitudes,  see e.g. \cite{brano} for a short review. In the special case of MHV amplitudes of ${\cal N}=4$ super-Yang-Mills theory the structure of (loop) amplitudes  simplifies sufficiently that can be described algebraically using Hopf algebras \cite{Duhr}. Recently, it has been observed that there exist relations between different amplitudes of that type given by the antipode map of the underlying Hopf algebra module structure \cite{Dixon,Liu:2022vck}. It would be interesting to understand these relations from the point of view of homotopy algebras and their morphisms.

\paragraph{Acknowledgments.}
We thank Athanasios Chatzistavrakidis for discussion and useful comments. The work is supported by the Croatian Science Foundation project IP-2019-04-4168.

\end{document}